\documentclass[preprint,5p,singlecolumn]{elsarticle}

\journal{Physics Letters B}

\usepackage{color}
\usepackage{amsfonts}
\usepackage{bm}
\usepackage{hyperref}
\usepackage{caption}
\usepackage{subcaption}
\usepackage{amsmath}
\usepackage{slashed} 
\usepackage{amssymb}
\usepackage{array}
\usepackage{longtable}
\usepackage[utf8]{inputenc}
\usepackage[dvipsnames]{xcolor}

\begin{document}

\begin{frontmatter}
\title{Top polarisation as a probe of CP-mixing top-Higgs coupling in $tjh$ signals}

\author{Riley Patrick\fnref{cor1}}
\ead{riley.patrick@adelaide.edu.au}
\author{Andre Scaffidi\fnref{cor2}}
\ead{andre.scaffidi@adelaide.edu.au}
\author{Pankaj Sharma\fnref{cor3}}
\ead{pankajs.phy@gmail.com}
 \fntext[cor1]{ORCID: 0000-0002-8770-0688}
 \fntext[cor2]{ORCID: 0000-0002-1203-6452}
 \fntext[cor3]{ORCID: 0000-0003-1873-1349}
\address{ARC Center of Excellence for Particle Physics at the Terascale, Department of Physics, University of Adelaide, 5005 Adelaide, South Australia}

\begin{abstract}
In this letter we explore beyond the Standard Model top-Higgs Yukawa couplings as a function of a CP-mixing parameter $\xi_t$ at the 14 TeV HL-LHC in the process $pp\to thj$. We observe that angular variables of the decay products of the top are non-trivially sensitive to $\xi_t$. This fact is exploited in a full detector level analysis that employs machine learning techniques to optimize signal sensitivity on a suite of variables, including lepton azimuthal angle. The key result of this study is an improved projected exclusion limit on $\xi_t$ even when including the realistic effects of detector smearing and a conservative estimate of systematic error.
\end{abstract}


\end{frontmatter}

\section{Introduction}
Since the discovery of a scalar particle at 125 GeV as predicted by the Standard Model (SM) at the Large Hadron Collider (LHC) in 2012 \cite{201230,20121} attention has turned to narrowing down it's properties. It is of great interest as to whether this particle behaves exactly as the SM predicts or if it is perhaps a beyond the SM (BSM) scalar that exists within an expanded Higgs sector.

The measurement of the Higgs couplings to fermions is an on-going area of research as many of these couplings can be small and require large luminosities to probe, which motivates the application of intelligent analysis techniques to probe the signal as optimally as possible. The Higgs coupling to the top is of particular interest as it is the largest, and so far the $t\bar{t}h$ coupling is probed via the loop-induced processes $hgg$ and $h\gamma\gamma$ which rely on decay rate measurements of the Higgs. It is well known that the production cross section of both $hgg$ and $h\gamma\gamma$  will be sensitive to a phase $\xi_t$ of the top Yukawa coupling \cite{Ren:2019xhp}. This phase mixes the CP-properties of the top-Higgs coupling and thus is CP violating. Assuming only the CP phase and keeping SM value for magnitude, strong constraints could be placed on the phase. However, those processes can only occur at one-loop level and beyond. Noting that new physics dynamics can contribute in the loop and thus may affect the accurate determination of top-Yukawa phase, the Higgs production in association with a top pair (dominant) and Higgs produced with a single top (sub-dominant) are the only means to directly probe this coupling. These processes have been established at above 5$\sigma$ in Refs~\cite{Aaboud:2018urx,Sirunyan:2018hoz}. It is found that increasing $|\xi_t|$ leads to a suppression of the $t\bar{t}h$ cross section and an enhancement of the $tjh$ cross section \cite{Chang:2014rfa,Kobakhidze2014}.
A study of this process has also been undertaken utilizing the Matrix-Element-Method (MEM) which shows that, with signal detection efficiencies on the order of a few percent, discovery could be made in the high luminosity phase of the LHC~\cite{Kraus:2019myc}.

The top quarks lifetime is so short that it decays before hadronization, meaning that its polarization information is preserved in the distribution of its decay products which can be measured directly by the detector, especially in its lepton angular distributions. In many new physics scenarios, it has been shown extensively in literature,  Refs.~\cite{Godbole2010,Godbole_2006,doi10106313327703,PhysRevD88074018,Huitu2011,Godbole2012,Rindani2011,RINDANI2012413,Rindani2013,Rindani2015},  that charged-lepton azimuthal distribution is a powerful probe of top quark polarisation in the lab frame. There are two advantages of studying the charged lepton azimuthal distribution: first it does not require reconstruction of top-rest frame which would need full information of top-quark momentum and second, it is unaffected by any new physics in the top-quark decay and thus making it an uncontaminated probe of top quark polarization. This variable is constructed by taking the azimuthal angle of the lepton decaying from the top with respect to the x-z plane, where the top quarks x-component is positive.

Top quark polarisation can be written in terms of $\xi_t$ \cite{Ellis2014,Yue:2014tya} and thus the decay products differential distributions will be effected by $\xi_t$ allowing for the improvement of analysis of this process. This has been exploited in a range of studies \cite{Ellis2014,Yue:2014tya,Rindani:2016scj}, however these have not been undertaken at detector level to provide more accurate reflections of achievable sensitivity. Furthermore, these studies have employed traditional cutflow methods rather than more modern and advanced machine learning (ML) techniques to optimize signal sensitivity. In this letter we for the first time employ a full detector level analysis to calculate the angular variables of the decay products of the process $pp\rightarrow thj$ with the optimization of signal sensitivity through ML algorithms.

This paper is structured as follows: section~\ref{sec:tophiggscoup} will cover the parametrization of the top-Higgs coupling and it's implementation, section~\ref{sec:sigback} will outline signal and backgrounds, section~\ref{sec:simreco} will outline event simulation and reconstruction, section~\ref{sec:results} will present the results of this analysis and finally we will conclude in section~\ref{sec:conc}.

\section{CP-mixed Top-Higgs Coupling}
\label{sec:tophiggscoup}
In this study, a CP-mixing parameter $\xi_t$ is introduced in the mass basis of the  top-Higgs sector via  the Lagrangian
\begin{align}
\label{eqn:Lag}
	\mathcal{L}_{tth} = -\frac{y_t}{\sqrt{2}} \bar{t}\left( \cos\xi_t + i\gamma_5\sin\xi_t\right){t}h\;.
\end{align}
The SM limit corresponds to when the mixing angle $\xi_t = 0$ and the Yukawa coupling adopts its SM value $y_t\rightarrow y_t^{SM} = \sqrt{2} m_{t} / v$, where $m_t$ is the mass of the top and $v\simeq 246$ GeV is the standard model Higgs vacuum expectation. We adopt a model independent approach in where the interaction Lagrangian in Eqn.~\ref{eqn:Lag} arises from an effective field theory (EFT) such as the dimention-6 opperators discussed in Refs.~\cite{AguilarSaavedra:2009mx,AguilarSaavedra:2008zc, Zhang:2012cd,Belusca-Maito:2015lna,He:2013tia}. We assume that the new physics scale $\Lambda$ of such an EFT is $\gtrsim$ TeV such that the mixing angle $\xi_t\in (-\pi, \pi]$ \cite{Harnik:2013aja}.  

Constraints on $y_t$ and $\xi_t$ from the $hgg$ and $h\gamma\gamma$ loop processes can be found in Refs~\cite{Ellis2014,Kobakhidze2014,Mileo2016,Nishiwaki2014,PhysRevD92015019,Cheung2013,PhysRevD90095009}. Further constraints including unitary violation in W and Z scattering with the top have been defined in Refs~\cite{PhysRevD.88.013014,PhysRevD.87.011702}. The strongest constraints come from precision electron dipole-moment (EDM) measurements \cite{Brod2013,Cirigliano:2016njn,Chien2016}, however these are done under assumptions which when relaxed allow for much looser constraints. Assuming a standard model value for $y_t$, collider constraints have rendered $\xi_t \in [0,2\pi/3]$ at 2$\sigma$ \cite{Mildner2016_1000065207}. For this study we also assume that $y_t$ and $WWh$ coupling adopt standard model values. Furthermore, for the sake of completeness, we perform the study on the entire region $\xi_t \in [0,\pi]$.

\section{Signal and Background}
\label{sec:sigback}
\begin{figure}[!ht]
\centering
\includegraphics[scale=0.50]{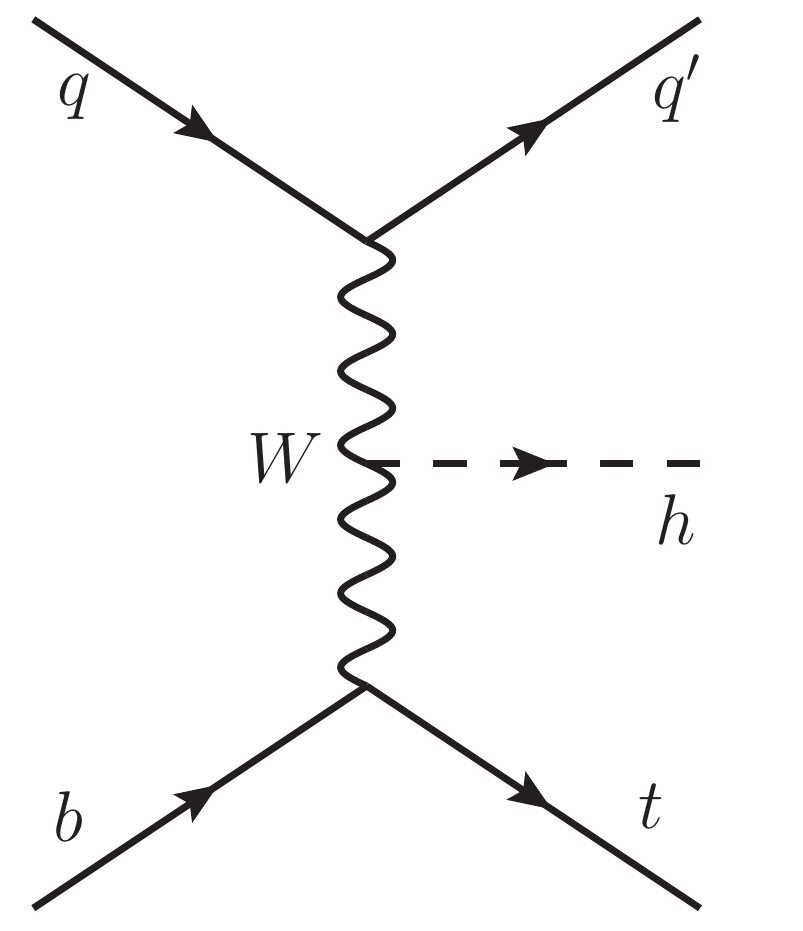}
\includegraphics[scale=0.50]{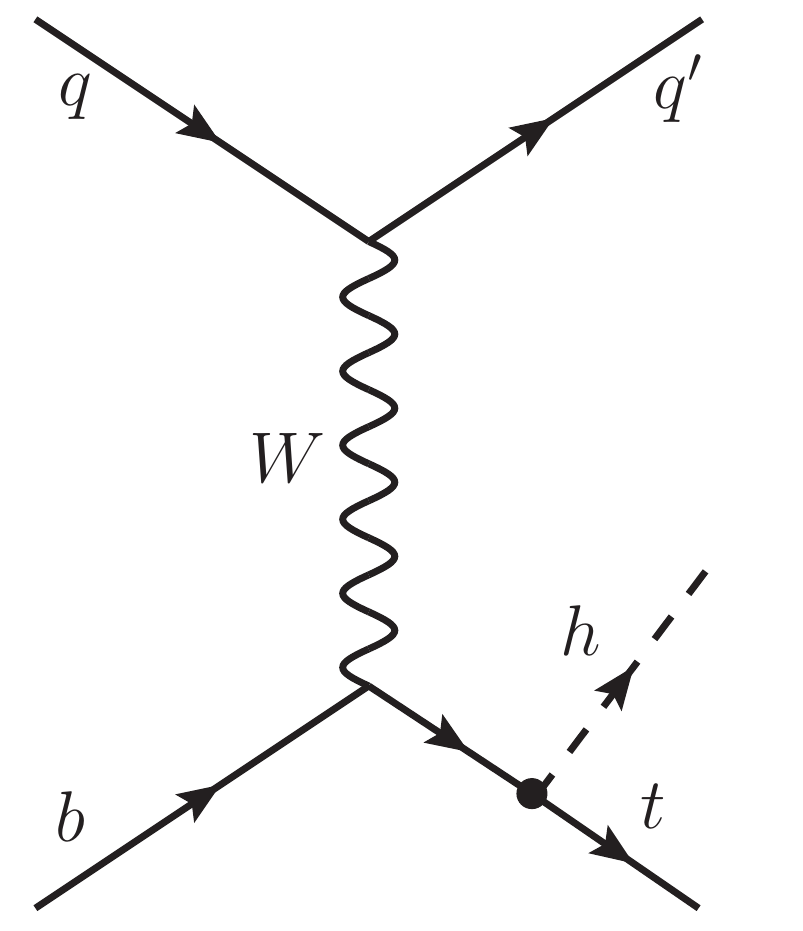}
\caption{\label{fig:prod-diag} Feynman diagrams for the dominant production process $bq\to thj$. }
\end{figure}

The process studied is Higgs production with an associated top quark and jet, $p p \to tjh$, at the 14 TeV LHC. Fig~\ref{fig:prod-diag} displays the dominant Feynman diagrams contributing to signal production. 

\begin{figure}[!ht]
\centering
\includegraphics[scale=0.19]{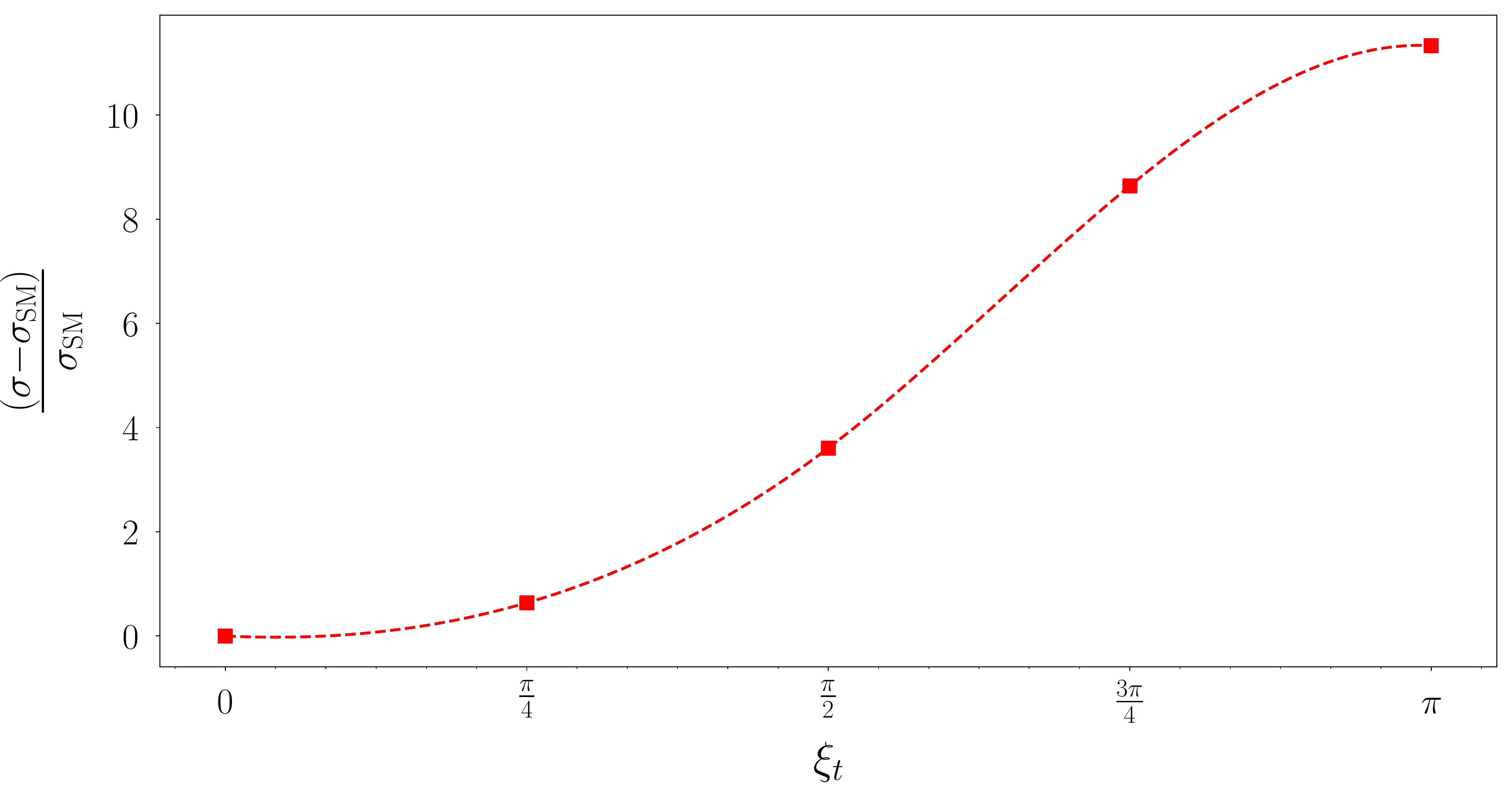}
\caption{\label{fig:xsections} The production cross section (red) and production cross section times $h\to\gamma\gamma$ branching ratio (blue) normalized with respect to standard model values for $\xi_t\in [0,\pi]$.}
\end{figure}

 Due to the extremely clean signature it provides, the decay mode of $h\to\gamma\gamma$ has comparable signal sensitivity to the $h\to b\bar{b}$ decay despite a much smaller branching ratio. We hence choose this decay mode of the Higgs for our analysis.
In Fig~\ref{fig:xsections} one can see the effect of the CP-mixing parameter $\xi_t$ on the production cross section of $tjh$. This effect is in agreement with previous results seen in Ref~\cite{Rindani:2016scj} and contains a maximum enhancement at $\xi_t = \pi$ of approximately 1200\%.

We demand a final state containing 1 or more $b$-jets, exactly 1 lepton (but not $\tau$) and at least 2 photons. The irreducible background for this process is $tj\gamma\gamma$ continuum. The sub-leading background is $t\bar{t}\gamma\gamma$ where a $b$-jet is mistagged as a light jet. Finally the background $Wjj\gamma\gamma$ where a light jet is mistagged as a $b$-jet exists, however it is found to be at least an order of magnitude lower than the previous two backgrounds~\cite{Wu:2014dba,Chang:2014rfa} and is ignored in this study.

\section{Event Simulation and Reconstruction}
\label{sec:simreco}

The parton level events are produced in \texttt{MG5\_aMC\_v2\_6\_0}~\cite{Frederix:2018nkq} then passed to \texttt{PYTHIA8}~\cite{Sjostrand:2007gs} for hadronization/fragmentation and finally \texttt{Delphes}~\cite{deFavereau:2013fsa} for detector effects.

We employ anti-kt jet clustering and take a $b$-tag efficiency of 77\%, a mistagging efficiency of 1\% and a lepton selection efficiency of 100\%. We also employ the following detector acceptance cuts:
\begin{align}
  p_T^{b,\ell} > 20 \text{GeV},~|\eta_{b,\ell}| < 2.5,~p^j_T > 25 \text{GeV},~|\eta_j| > 2.5
\end{align}
The cut on $|\eta_j|$ is employed to take advantage of the forwardness of the light jet which is characteristic of the $tjh$ signal.

As we are selecting exactly 1 lepton we are able to calculate the longitudinal momentum of the neutrino decaying from the top. We do this using the following quadratic equation:
\begin{equation}
p_\nu^z = \frac{1}{2p_{\ell T}^2}\left(A_Wp^z_{\ell}\pm E_\ell\sqrt{A_W^2\pm 4p^2_{\ell T}E^2_{\nu T}}\right),
\end{equation}
 where, $A_W = M^2_{W^{\pm}} + 2p_{\ell T}\cdot E_{\nu T}$. We choose the solution for $p_\nu^z$ that is real and that when combined with the remainder of the neutrino and lepton 4-vector components produces an invariant mass closest to the $W$ boson mass. After this is done the top quark is reconstructed from the neutrino, the lepton and the $b$-jet which best reproduces the top quark invariant mass.

\begin{figure}[ht!]
\centering
\begin{subfigure}{0.5\textwidth}
\centering
\includegraphics[scale=0.2]{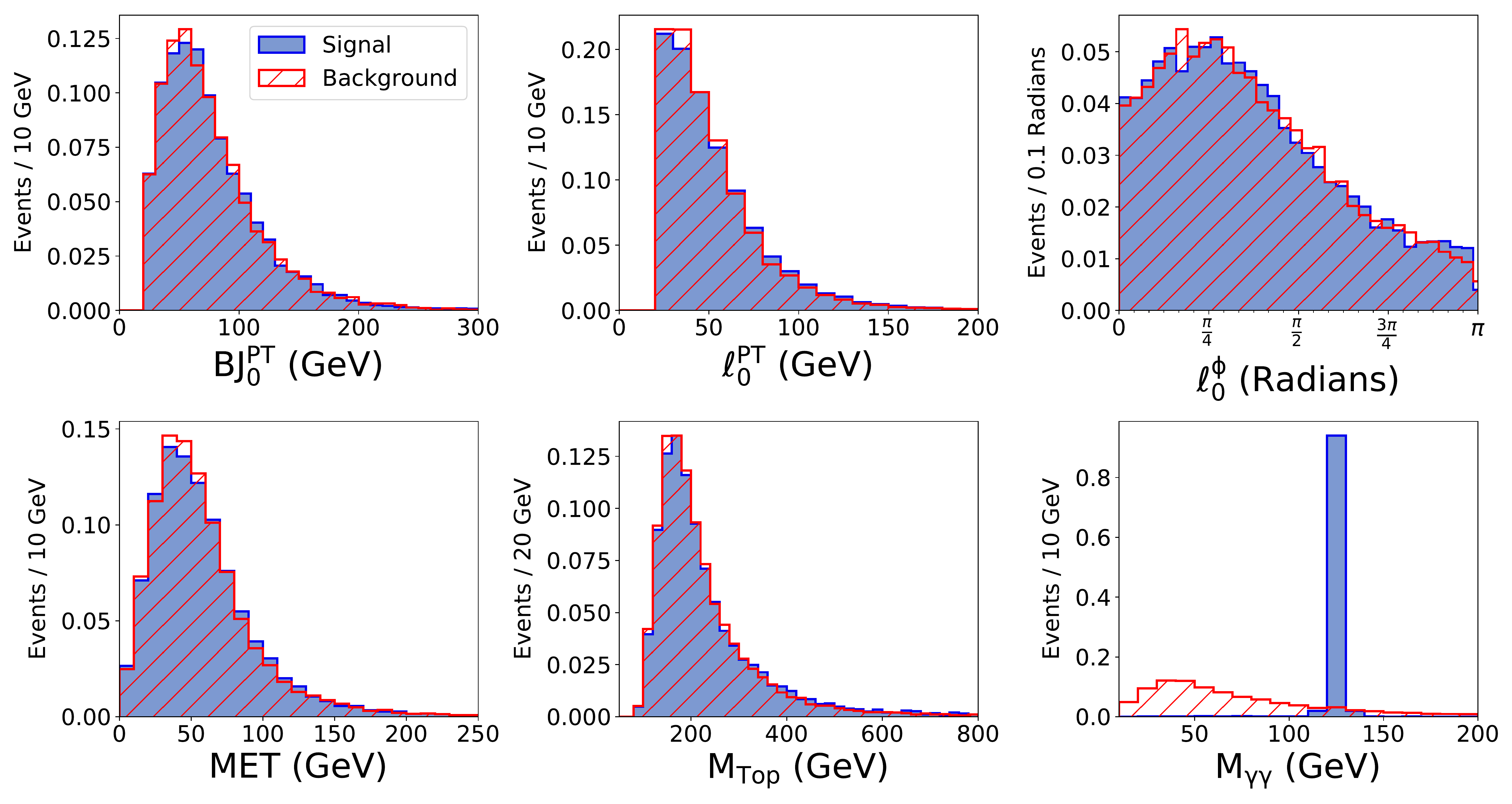}
\caption{}
\end{subfigure}
\\
\begin{subfigure}{0.5\textwidth}
\centering
\includegraphics[scale=0.2]{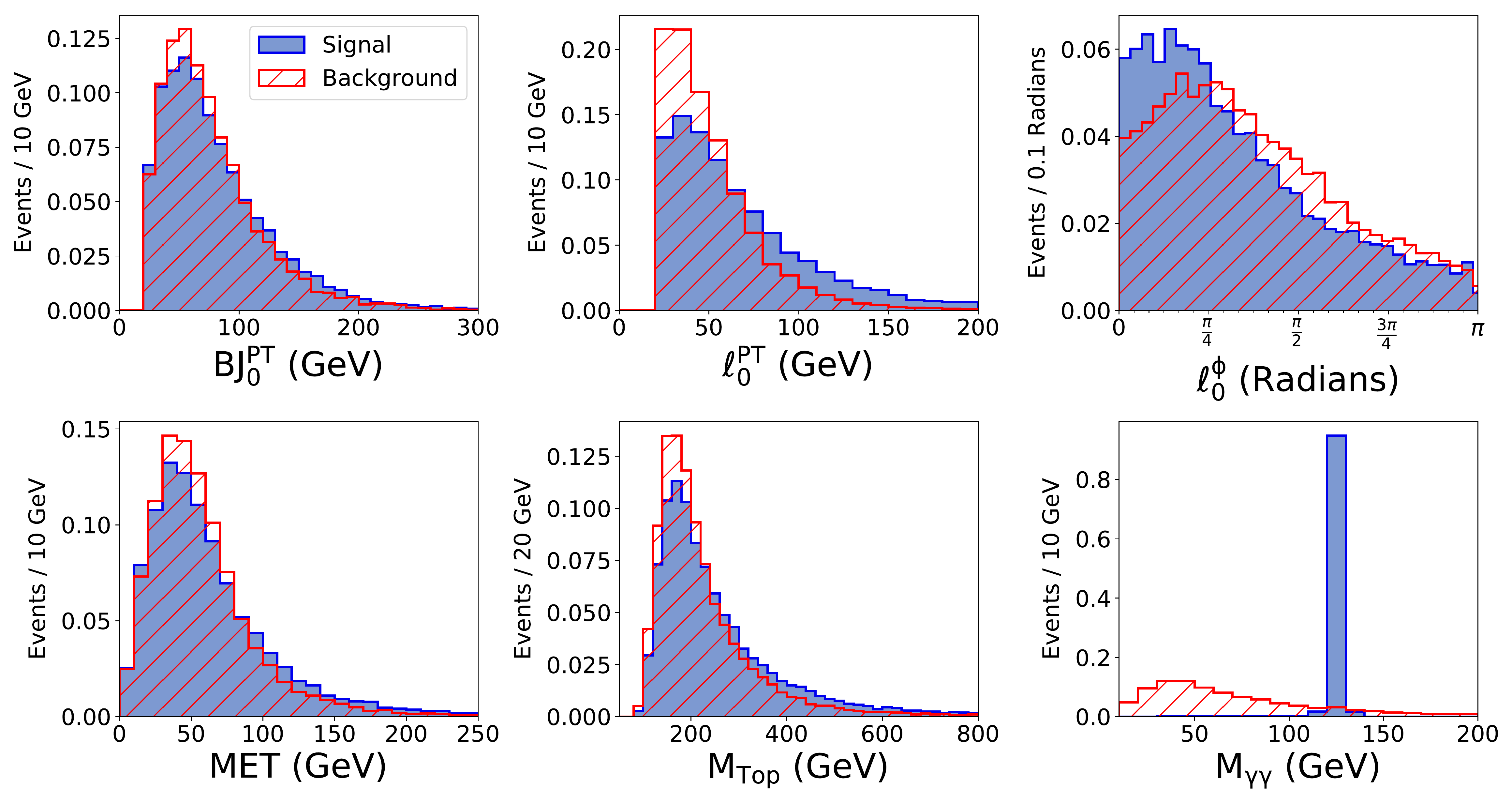}
\caption{}
\end{subfigure}
\caption{\label{fig:variables} Examples of variables employed in the analysis for the $\xi = 0$ (a) and $\xi = \frac{\pi}{2}$ (b) benchmarks.}
\end{figure}

In Fig~\ref{fig:variables} we present the variables selected for the numerical analysis to come. In the $\ell^\phi_0$ plot (upper right of each subfigure) the lepton azimuthal distribution generated from the prescription above can be seen for the hardest lepton in each event. It is clear that the value of $\xi_t$ is impacting this distribution significantly as the SM value of $\xi_t = 0$ presents a distribution identical to the background while the fully CP-odd value of $\xi_t = \frac{\pi}{2}$ presents a far more skewed distribution. In addition, the variables show that the reconstruction is faithfully producing invariant mass distributions for the top quark and Higgs.

\section{Results}
\label{sec:results}

\begin{figure}[!ht]
\centering
\includegraphics[scale=0.19]{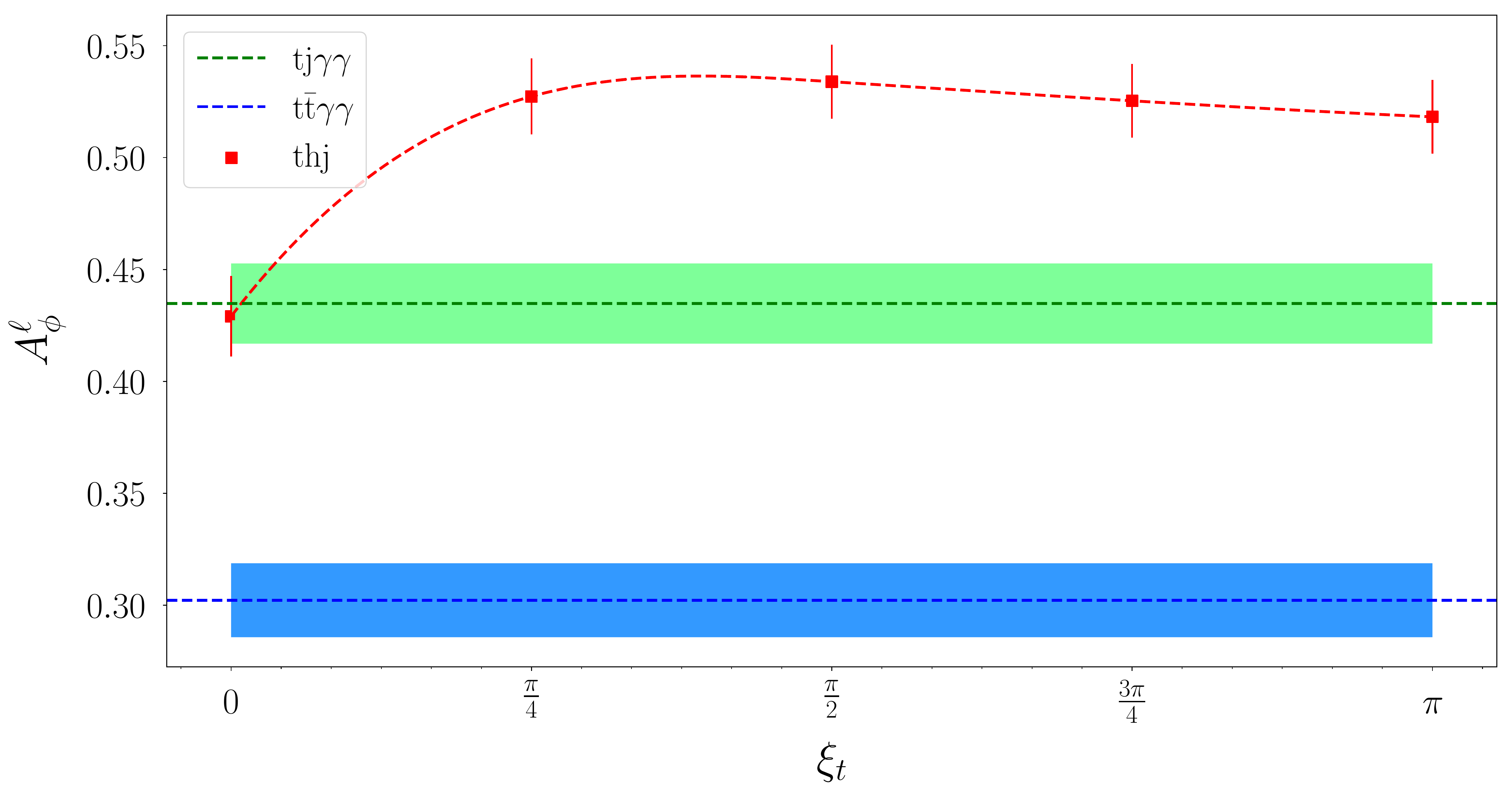}
\caption{\label{fig:asymmetry_phi_lep} The $A^{\ell}_{\phi}$ asymmetry as a function of $\xi_t$ for the signal and background.}
\end{figure}

As seen in Ref~\cite{Rindani:2016scj}, we can construct the lab frame left-right asymmetry of the charged lepton using:
\begin{align}
A_{\phi}^{\ell} = \frac{ \sigma\left(\cos\phi > 0\right) - \sigma\left(\cos\phi < 0\right) }
					   { \sigma\left(\cos\phi > 0\right) + \sigma\left(\cos\phi < 0\right)}
\end{align}
Fig~\ref{fig:asymmetry_phi_lep} displays this asymmetry as a function of $\xi_t$ which takes a maximimum at $\xi_t = \frac{\pi}{2}$. However when compared to the parton level calculation of this asymmetry found in Ref~\cite{Rindani:2016scj} it can be seen that the detector effects lead to a flattening of this curve.

Standard cut flows for this signal have been performed in the past such as in Refs~\cite{Yue:2014tya}\cite{Rindani:2016scj} and thus we do not perform one in this work. Instead we employ a boosted decision tree analysis on the variables found in Fig~\ref{fig:variables} using the \texttt{Toolkit for Multivariate Data Analysis} (TMVA)~\cite{Harnik:2013aja}.

\begin{table}[!ht]
\centering
\begin{tabular}{|lcccccc|}
\hline
$\xi_t$ & $N^b_s$ & Cut  & $N^a_s$ & $N^a_b$ & $Z$ & $Z_{0.2}$ \\\hline
$0$ & 8 & 0.1146  & 7 & 21 & 1.28 & 1.04 \\\hline
$\frac{\pi}{4}$ & 23 & 0.0910  & 20 & 24 & 3.05 & 2.44 \\\hline
$\frac{\pi}{2}$ & 89 & 0.0819  & 83 & 27 & 7.92 & 7.04 \\\hline
$\frac{3\pi}{4}$ & 198 & 0.0317  & 191 & 35 & 12.70 & 11.52 \\\hline
$\pi$ & 255 & 0.0503 & 244 & 34 & 14.64 & 13.55 \\\hline
\end{tabular}
\caption{\label{tab:significance} \\ Table of signal sensitivity defined as: $Z = S/\sqrt{S+B+\left(\Delta B\right)^2}$ after applying the optimized cut generated by BDT analysis. A luminosity of 3000fb$^{-1}$ is chosen and the number of background events before cuts is $N^b_b = 1076$.}
\end{table}

Tab~\ref{tab:significance} presents the BDT results for each value of $\xi$. The column labels are as follows: $N^b_s$ is the number of signal events before cuts while $N^a_s$ and $N^a_b$ are the number of signal and background events after cuts respectively. The number of background events before cuts was $N^b_b = 1076$. The column labelled ``Cut'' is the position of the optimal cut on the BDT classifier distribution. $\sigma$ is the signal sensitivity considering no systematic error, while $\sigma_{0.2}$ is the signal sensitivity with a flat 20\% systematic error. The systematic error of 20\% was taken as an estimate of the overall level of systematic error in typical 1-lepton plus jets final state experiments~\cite{Aaboud:2017aeu}. The results presented in this table show, as expected, high values of $\xi_t \geq \frac{\pi}{2}$ are strongly inconsistent with background only. The 95$\%$ C.L. exclusion of approximately $\xi_t = 0.54$ without systematic and $\xi_t = 0.68$ with systematic is expected for 3ab$^{-1}$ of data at the HL-LHC, a significant improvement on the $\xi_t = 0.79$ exclusion placed in Ref~\cite{Rindani:2016scj}. This is due to a combination of the BDT analysis and the additional top polarisation variable included. It is likely that had this analysis been undertaken at parton level like previous analyses of this process then improvements on the constraints would be even larger, that is to say that detector effects have likely reduced the overall positive impact of the BDT analysis and top polarisation variable.

\section{Conclusion}
\label{sec:conc}

The direct detection of top-Higgs coupling has now been achieved via the process $pp \to t\bar{t}h$, however the properties of this coupling still require further study. The process $pp \to tjh$ provides a good window into the charge-parity properties of the coupling as increasing values of the top-Higgs coupling phase $\xi_t$ lead to increased cross sections, while the $t\bar{t}h$ process experiences decreases.

In this work we have introduced a CP-mixing parameter $\xi_t$ to the SM top-Higgs coupling via an effective operator. We have explored the well studied effects of this variable on top and jet associated Higgs production. Previous studies were then expanded on by performing a full detector level analysis of this process including the variable defined from the azimuthal distribution of the lepton decaying from the top which provides a powerful insight into top polarisation. Results were then further improved via the application of a ML algorithm, namely boosted decision tree analysis, to optimize signal sensitivity.

The key result of this study is a projected 95$\%$ median exclusion of $\xi_t \leq 0.54$ when not considering systematic errors and $\xi_t \leq 0.68$ when considering a conservative level of systematic error with 3ab$^{-1}$ of luminosity, a significant improvement over previous analyses of this process. It is reasonable that the HL-LHC can provide very strong limits on $\xi_t$ even in pessimistic scenarios.

\section*{Acknowledgements}
This work is supported by the University of Adelaide and the Australian Research Council through the ARC Center of Excellence for Particle Physics (CoEPP) at the Terascale (grant no. CE110001004).

\bibliographystyle{elsarticle-num}
\bibliography{tqh}

\end{document}